\documentclass[prl, twocolumn,showpacs,amsmath,floatfix,superscriptaddress, showkeys]{revtex4-1}
\usepackage[colorlinks,linkcolor=blue,citecolor=blue,breaklinks]{hyperref} 
\usepackage[hyphenbreaks]{breakurl}
\usepackage{CJK}
\usepackage{graphicx,color,epstopdf}
\usepackage{mathptmx}
\usepackage{dcolumn}
\usepackage{makecell}
\usepackage{amssymb,bm,mathrsfs}
\usepackage{txfonts}
\usepackage{dcolumn,multirow}
\usepackage{slashed,indentfirst}
\usepackage{comment}
\usepackage{float} 
\usepackage{ulem}
\usepackage{color}

\newcommand{\delete}{\bgroup\markoverwith{\textcolor{blue}{\rule[0.5ex]{2pt}{1pt}}}\ULon}

\begin{document}
\begin{CJK}{UTF8}{gkai}
\title{Towards a Unified Description of Isoscalar Giant Monopole Resonances in a Self-Consistent Quasiparticle-Vibration Coupling Approach}
\author{Z.~Z.~Li}
\affiliation{School of Nuclear Science and Technology, Lanzhou University, Lanzhou 730000, China}
\affiliation{Frontiers Science Center for Rare isotope, Lanzhou University, Lanzhou 730000, China}
\affiliation{Dipartimento di Fisica, Universit\`a degli Studi di Milano, via Celoria 16, 20133 Milano, Italy}
\author{Y.~F.~Niu}\email{niuyf@lzu.edu.cn}
\affiliation{School of Nuclear Science and Technology, Lanzhou University, Lanzhou 730000, China}
\affiliation{Frontiers Science Center for Rare isotope, Lanzhou University, Lanzhou 730000, China}
\author{G.~Col\`o}\email{gianluca.colo.mi.infn.it}
\affiliation{Dipartimento di Fisica, Universit\`a degli Studi di Milano, via Celoria 16, 20133 Milano, Italy}
\affiliation{INFN sezione di Milano, via Celoria 16, 20133, Milano, Italy}

\begin{abstract}

``Why is the EoS for tin so soft?" is a longstanding question, which prevents us from determining the nuclear incompressibility $K_\infty$ accurately. To solve this puzzle, a fully self-consistent quasiparticle random phase approximation (QRPA) plus quasiparticle-vibration coupling (QPVC) approach based on Skyrme-Hartree-Fock-Bogoliubov is developed. We show that the many-body correlations introduced by QPVC, which shift the ISGMR energy in Sn isotopes by about 0.4 MeV more than the energy in $^{208}$Pb, play a crucial role in providing a unified description of the ISGMR in Sn and Pb isotopes. The best description of the experimental strength functions is given by SV-K226 and KDE0, which are characterized by incompressibility values $K_\infty=$ 226 MeV and 229 MeV, respectively, at mean field level.

\end{abstract}
\maketitle
\end{CJK}
The nuclear compression-mode resonances, especially the isoscalar giant monopole resonance (ISGMR), have been suggested to be a unique probe for the nuclear incompressibility $K_\infty$, which is important for constraining the nuclear equation of state (EoS) \cite{Harakeh_2001_book_GRs, Blaizot_1980_PhysRep_ISGMR, Garg_2018_PPNP_K0}, and, in turn, impacts our understanding of the astrophysical processes, such as supernova explosion and neutron star merging \cite{Oertel_2017_RMP_EoS, Perego_2022_PRL}. However, as we discuss in what follows, it is still unclear
how to get a unified description of ISGMR strength functions in different isotopic chains, and this prevents us from determining the nuclear incompressibility accurately.
 
In $^{208}$Pb, the ISGMR was measured in the Research Center for Nuclear Physics (RCNP) by means of inelastic $\alpha$-scattering  \cite{Patel_2013_PLB, Uchida_2004_PRC_ISGMR} and inelastic deuteron scattering \cite{Patel_2014_PLB_Pb208_Dscattering}, and in Texas A\&M University (TAMU) by means of inelastic $\alpha$-scattering \cite{Youngblood_1999_PRL_ISGMR, Youngblood_2004_PRC_ISGMR}. These measurements, together with those performed in $^{90}$Zr \cite{Krishichayan_2015_PRC_Zr, Uchida_2004_PRC_ISGMR,Gupta_2016_PLB_Zr}, hint to a value $K_\infty=240\pm20$ MeV for the nuclear incompressibility \cite{Garg_2018_PPNP_K0}. However, it was found that in even-even stable tin isotopes, $^{112-124}$Sn, the ISGMR centroid energy is overestimated (by about 1 MeV) by the same models which reproduce the ISGMR centroid energy well in $^{208}$Pb \cite{T.Li_2007_PRL, T.Li_2010_PRC_Sn}. This means that the incompressibility value that is deduced from Sn is lower than the one deduced from Pb or, in other words, the EoS is softer \cite{Piekarewicz_2007_PRC_Soft, Piekarewicz_2010_JPG, Garg_2007_NPA_fluffy}. Later, 
a similar ``softness'' was also found in even-even $^{106,110-116}$Cd isotopes \cite{Patel_2012_PLB_Cd} and even-even $^{94-100}$Mo isotopes \cite{Howard_2020_PLB_Mo}.

In the present work, we want to attack the question ``Why is the EoS for tin so soft?'' in a novel manner. In the papers we have mentioned, the 
correspondence between $K_\infty$ and the ISGMR energy is established by means of QRPA calculations. QRPA is a well-known microscopic method to describe the Giant Resonances (GRs). The axial deformation within the QRPA model helps to explain the softness in Mo isotopes  \cite{Colo_2020_PLB_Mo}, but tin nuclei are spherical. A lot of efforts have been devoted to exploring the pairing effects within the self-consistent QRPA model \cite{J.Li_2008_PRC, L.G.Cao_2012_PRC_CdSn, Avogadro_2013_PRC_Softness}. It has been shown that surface pairing can partly reconcile tin and lead results (compared to volume pairing and mixed pairing) \cite{J.Li_2008_PRC, L.G.Cao_2012_PRC_CdSn}. However, there is no strong argument on which type of pairing force should be favored over others. Moreover, the proposed ``mutually enhanced magicity'' (MEM) effect on nuclear incompressibility, that is another attempt to solve the Sn/Pb puzzle, has also been ruled out by the measurement in $^{204,206}$Pb \cite{Patel_2013_PLB, Khan_2009_PRC_MEM}. 

Despite the big successes achieved in QRPA, the widths, decay properties, and fine structures of GRs are not well described in QRPA because of the lack of coupling with more complex configurations than the two-quasiparticle ones. The particle-vibration coupling (PVC) effects have been proven to be crucial for reproducing the widths \cite{Bertsch_1983_RMP_Damping} and describing the decay properties of GRs \cite{W.L.Lv_2021_PRC_GRs}. These particle-vibration coupling effects have been included in a self-consistent way, based on both nonrelativistic density functional theory 
\cite{Lyutorovich_2012_PRL,Lyutorovich_2015_PLB,Tselyaev_2016_PRC, Roca-Maza_2017_JPG_PVC, Y.F.Niu_2016_PRC_QPVC} and relativistic density functional theory \cite{Litvinova_2007_PRC_PVC, Litvinova_2007_PLB_RPVC, Litvinova_2008_PRC_QPVC}. Although many works have been devoted, e.g., to the electric dipole resonances \cite{Lyutorovich_2012_PRL, Egorova_2016_PRC_QTBA, Avdeenkov_2011_PRC_QPVC_E1, Litvinova_2008_PRC_QPVC}, within the quasiparticle-vibration coupling framework, this is not the case for the ISGMR.

In this Letter, a fully self-consistent Quasiparticle Random Phase Approximation plus Quasiparticle-Vibration Coupling model (QPVC) is developed, based on the Skyrme-Hartree-Fock-Bogoliubov (SHFB) framework. In it, we consider both QPVC effects and pairing effects self-consistently. We give a short description of our framework, and more details will be discussed in a further publication. We aim to demonstrate that the consideration of QPVC effects is crucial in order to reach a unified description of the ISGMR energies in Ca, Sn, and Pb isotopes. 

\begin{figure*}
	\centering
	\includegraphics[width=0.99\textwidth]{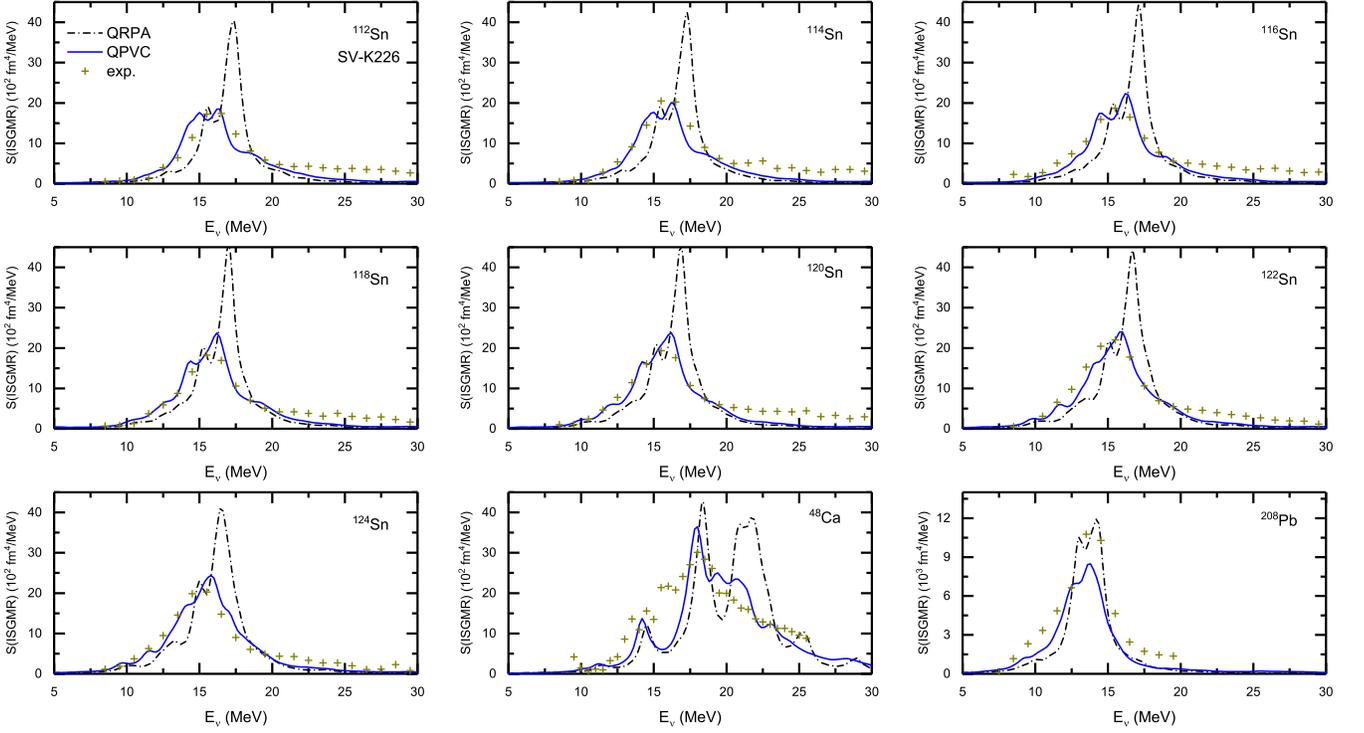}
	\caption{(Color online) ISGMR strength functions in even-even $^{112\textnormal{-}124}$Sn, $^{48}$Ca, and $^{208}$Pb isotopes, calculated either by (Q)RPA using a smoothing with Lorentzian having a width of 1 MeV (dash dot [black] line), or (Q)RPA+(Q)PVC (solid [blue] line). The SV-K226 Skyrme force is used. The experimental data are given by green crosses \cite{T.Li_2007_PRL, Patel_2013_PLB, Olorunfunmi_2022_PRC}.} \label{fig-1}
\end{figure*} 

We start from the spherical SHFB code in the coordinate space from Ref. \cite{Bennaceur_2005_CPC_SHFB}. The so-called volume pairing force, $v_{pp}(\pmb r_1, \pmb r_2) = V_{0,q} \delta (\pmb r_1 - \pmb r_2)$, is used to describe the pairing interaction, where $q$ labels either neutrons or protons. $V_{0,q}$ is adjusted by fitting the neutron (proton) pairing gaps according to the five-point formula \cite{Bender_2000_EPJA_PGaps}. The SHFB solution is performed in a spherical box 20 fm, with a mesh of 0.1 fm. The QRPA equations are solved in the canonical basis. The spurious state, caused by the violation of particle number, is removed as in Ref. \cite{J.Li_2008_PRC}. For the convergence of the results, a single-particle energy cutoff $E_\textnormal{cut}=60$ MeV, and an angular momentum cutoff $j_\textnormal{max}=15/2$ are set for Ca isotopes, while larger cutoffs $E_\textnormal{cut}=100$ MeV, $j_\textnormal{max}=21/2$ are used for Sn and Pb isotopes. On top of QRPA, we have included the coupling with phonons having $J^\pi=$ $0^+$, $1^-$, $2^+$, $3^-$, $4^+$, $5^-$, with energy less than 30 MeV and exhausting a fraction of non-energy-weighted (isoscalar or isovector) sum rule larger than 2\%. The subtraction procedure is adopted, as described in \cite{Tselyaev_2007_PRC_QTBA}.

The sum rules, or $k$-th moments of the strength function $S(E)$ are defined as $m_k=\int_0^\infty S(E) E^k dE$. In our case, $S(E)$ is with respect to the operator $\hat F_{00} = \sum_{i=1}^A r_i^2$. The fulfillment of the energy-weighted sum rule (EWSR) $m_1$ (\%), and inverse energy-weighted sum rule (IEWSR) $m_{-1}$ (fm$^4$/MeV), calculated by QRPA+QPVC, have been checked. Taking $^{120}$Sn with the SV-K226 Skyrme set as an example, the EWSR is given as 215185.4 fm$^{4}$MeV from the expectation value of the double-commutator on the nuclear ground state. In a fully self-consistent PVC approach, $m_1$ should be fulfilled in the case without subtraction \cite{S.H.Shen_2020_PRC_PVC}. Up to 100 MeV, $m_{1}$ is indeed exhausted at 98.7\%. With the subtraction procedure, $m_{1}$ is exhausted by $106.8\%$, while $m_{-1}$ is $818.45$ fm$^4/$MeV, which is nearly equal to the one in QRPA (820.71 fm$^4/$MeV) as discussed in \cite{Roca-Maza_2017_JPG_PVC}. 

There are many choices of characteristic energy for GRs, such as the centroid energy $m_1/m_0$, the constrained energy $\sqrt{m_1/m_{-1}}$, and the scaling energy $\sqrt{m_3/m_1}$. In the following, we will use the constrained energy $\sqrt{m_1/m_{-1}}$ for our discussion since $m_{-1}$ is unchanged in the case of QPVC with subtraction. Our conclusions would remain the same if we were to choose another definition for the ISGMR energy. The ISGMR energies are calculated in the energy interval 10--30 MeV for Ca, and 5--25 MeV for Sn and Pb, because the strength is negligible outside these intervals.


In Fig. \ref{fig-1}, we show the strength functions of the ISGMR, obtained either in the framework of (Q)RPA by using a smoothing with Lorentzian having a width of 1 MeV (dash dot [black] line), or within (Q)RPA+(Q)PVC (solid [blue] line), using the SV-K226 Skyrme force, in the even-even 
$^{112\textnormal{-}124}$Sn, $^{48}$Ca, and $^{208}$Pb nuclei. We compare the results with the experimental ones ([green] crosses) \cite{T.Li_2007_PRL, Patel_2013_PLB, Olorunfunmi_2022_PRC}. In general, with the inclusion of (Q)PVC effects, the results are significantly improved with respect to (Q)RPA, so we can achieve a good description of data both in the light $^{48}$Ca isotope, medium-heavy Sn isotopes, and heavy $^{208}$Pb. In $^{112-124}$Sn, QRPA gives one small peak and one higher peak while the experimental strength displays a broad single peak. The ISGMR energies are higher than the experimental ones, as pointed out in previous papers \cite{Piekarewicz_2007_PRC_Soft, Garg_2007_NPA_fluffy}. With the inclusion of QPVC effects, widths are comparable with the experimental ones (cf. also \cite{Tselyaev_2009_PRC_Sn-ISGMR}). Moreover, within the self-consistent QRPA+QPVC model, the downward shifts of energies by $0.7$--$0.8$ MeV  
(with respect to QRPA) make the ISGMR energies in agreement with data, along the whole Sn isotopic chain. In the case of $^{48}$Ca, the strength function has two main peaks in the RPA calculation, while the experimental strength shows only a single main peak. With PVC effects, the first peak is 
slightly moved to lower energy, and the second peak becomes fragmented, making the results closer to experiment. In the case of $^{208}$Pb, the ISGMR energy given by RPA agrees with the experimental one. With PVC effects, the ISGMR has a very small shift to lower energy, and a larger width that again makes the result closer to the experiment finding.

\begin{figure}
	\centering
	\includegraphics[width=0.45\textwidth]{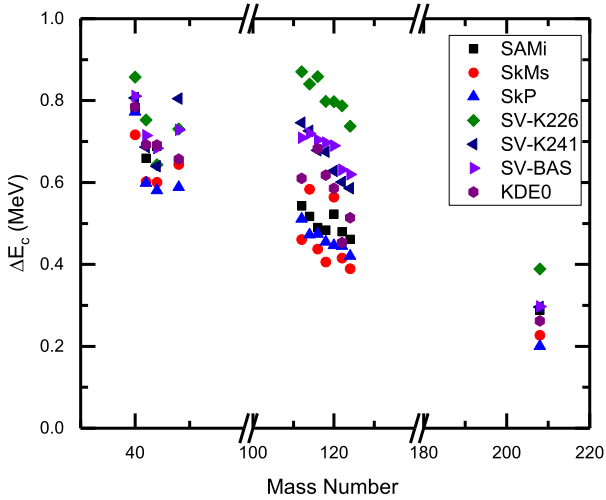}
	\caption{(Color online) The energy shifts of ISGMR from (Q)RPA to (Q)RPA+(Q)PVC ($E_c=\sqrt{m_1/m_{-1}}$) 
	in even-even $^{40\textnormal{-}44,48}$Ca, $^{112-124}$Sn, and $^{208}$Pb isotopes with 7 Skyrme sets: SAMi (black square), SkMs (red circle), SkP (up blue triangle), SV-K226 (olive diamond), SV-K241 (left navy triangle), SV-BAS (right violet triangle), and KDE0 (purple hexagon).} \label{fig-2}
\end{figure}

In Fig. \ref{fig-2}, the energy shifts of the ISGMR from (Q)RPA to (Q)RPA+(Q)PVC (considering $E_c=\sqrt{m_1/m_{-1}}$) are given in even-even $^{40\textnormal{-}44,48}$Ca, $^{112-124}$Sn and $^{208}$Pb isotopes using 7 Skyrme parameter sets: SAMi (black square), SkM* (red circle), 
SkP (up blue triangle), SV-K226 (olive diamond), SV-K241 (left navy triangle), SV-BAS (right violet triangle), and KDE0 (purple hexagon). These forces are obtained by different groups, using quite different fitting protocols, and span a large range of $K_\infty$ from $201$ MeV to $245$ MeV. As shown in Fig. \ref{fig-2}, the energy shifts from (Q)RPA to (Q)RPA+(Q)PVC are less than $1$ MeV in general. In detail, the results depend on the Skyrme set that is used, and the associated dispersion in the Sn isotopes is about 0.5 MeV. However, from Ca, Sn, to Pb, the energy shifts become smaller. The energy shifts in Ca and Sn isotopes are about 0.4 MeV larger than the ones in $^{208}$Pb, and this makes it possible to describe well the ISGMR in these different isotopes at the same time. 

\begin{figure}
	\centering
	\includegraphics[width=0.45\textwidth]{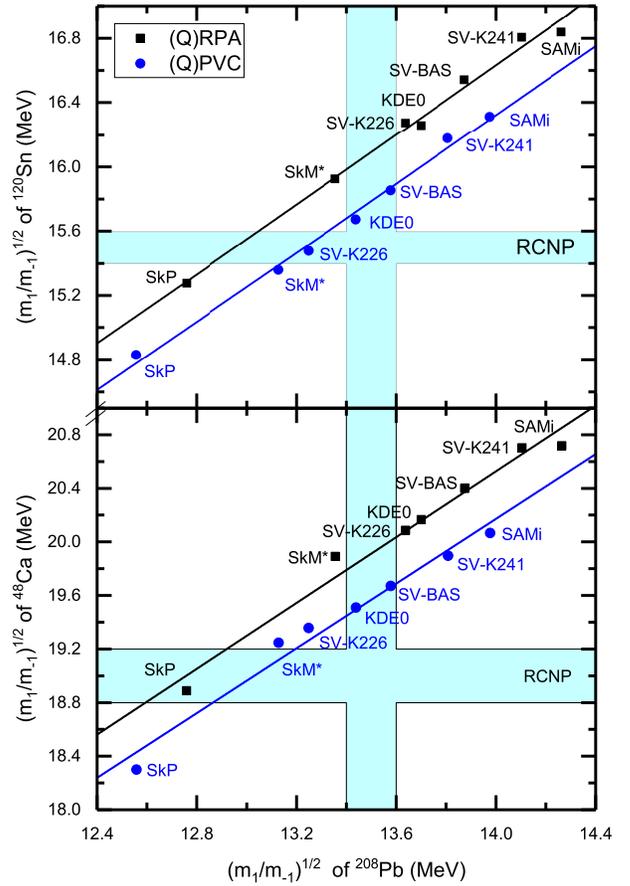}
	\caption{(Color online) The ISGMR energies in $^{208}$Pb vs. the ones in $^{120}$Sn (upper panel), 	and $^{48}$Ca (lower panel). These are calculated by (Q)RPA (black square), and by (Q)RPA+(Q)PVC (blue circle) using 7 different Skyrme parameters. The regression lines are obtained by a least-square linear fit of the (Q)RPA results and (Q)RPA+(Q)PVC results, respectively. The experimental data and their uncertainties, taken from \cite{T.Li_2007_PRL, Patel_2013_PLB, Olorunfunmi_2022_PRC}, are displayed by means of cyan-colored bands.} \label{fig-3}
\end{figure}

The linear correlation between the ISGMR energies and $K_\infty$, calculated by different models, is often used as a way to constrain $K_\infty$ \cite{Blaizot_1995_NPA}. Therefore, in principle, one can expect a linear correlation of the ISGMR energy in different nuclei. In the upper panel of Fig. \ref{fig-3}, we show that there is a linear correlation between the ISGMR energies in $^{120}$Sn and $^{208}$Pb, both in the (Q)RPA case (black square) and in the (Q)PVC case (blue circle). We use the 7 Skyrme interactions that we have already mentioned. The regression lines are obtained by a least-square fit. For the calculations in $^{120}$Sn, the volume pairing interaction is adopted. In the lower panel of Fig.  \ref{fig-3}, the same kind of linear correlation, between the ISGMR energies in $^{48}$Ca and $^{208}$Pb, are presented. The experimental data and their uncertainties are taken from \cite{T.Li_2007_PRL, Patel_2013_PLB, Olorunfunmi_2022_PRC}. The interesting point is that one cannot describe the ISGMR energy of $^{120}$Sn (or $^{48}$Ca) and $^{208}$Pb simultaneously at QRPA level, since the regression line is far from the crossing zone of experimental bands for these two nuclei. For example, SV-K226 and KDE0 give a good description in $^{208}$Pb, but they overestimate the ISGMR energy in $^{120}$Sn by about 0.8 MeV. However, with the consideration of (Q)PVC effects, the regression line given by the QPVC results is shifted downwards by about 0.3 MeV, so it is marginally compatible with the experimental zone. In this case, SV-K226 and KDE0 give pretty good ISGMR energies in both $^{120}$Sn and $^{208}$Pb. The same conclusion is achieved looking at $^{48}$Ca and $^{208}$Pb, as shown in the lower panel. In the latter case, PVC effects play a unique role since there is no pairing. In summary, our results suggest the (Q)PVC effects are crucial in order to reach a unified description of the ISGMR in Ca, Sn, and Pb isotopes at the same time.

The constraint on $K_\infty$ are less clear in the theories beyond mean field, because it is hard to calculate $K_\infty$ due to the ultraviolet divergence associated with zero-range effective interactions \cite{Moghrabi_2010_PRL_BMF_Ultraviolet}. Nevertheless, with the subtraction procedure, we can assume that $K_\infty$ will be the same as that at the mean field level. The deviations of ISGMR energies from experimental data [$|E_c^\textnormal{theo.}-E_c^\textnormal{exp.}|$ (MeV)] in $^{48}$Ca, $^{120}$Sn, and $^{208}$Pb are given in Tab. \ref{tab-1}, calculated by (Q)RPA and (Q)RPA+(Q)PVC using SkP, SkM*, SV-K226, KDE0, SV-bas, SV-K241, and SAMi. The Table shows that, at (Q)RPA level, $^{48}$Ca and  $^{120}$Sn prefer SkP with a small incompressibility, $K_\infty=201$ MeV. However, $^{208}$Pb prefers SkM*, SV-K226, and KDE0, with $K_{\infty}$ ranging from $218$ to $229$ MeV. With the inclusion of (Q)PVC effects, $^{48}$Ca prefers SkM* and SV-K226, $^{120}$Sn prefers SkM*, SV-K226, and KDE0, while $^{208}$Pb prefers SV-K226, KDE0, and SV-bas. Thus, SV-K226 and KDE0 describe all three nuclei very well at the same time, with $K_{\infty} = $ $226$ MeV and $229$ MeV respectively: this is consistent with the constraint $240\pm20$ MeV, obtained previously from the ISGMR of $^{208}$Pb 
in QRPA \cite{Garg_2018_PPNP_K0}.
 
\begin{table}
	\begin{tabular}{c ccccccc} \hline \hline  
		&   SkP     & SkM*      &  SV-K  & KDE0   & SV-bas  & SV-K  & SAMi   \\  
		$K_\infty$ 	   
		&   201     & 217       &  226      &  229   & 233     & 241      & 245     \\ \hline  
		(Q)RPA &  & & &  &&& \\  
		$^{48}$Ca   &  0.11     & 0.89      &  1.09     & 1.17   & 1.40    & 1.70     & 1.72    \\  
		$^{120}$Sn  &  0.22     & 0.43      &  0.78     & 0.76   & 1.05    & 1.31     & 1.34    \\  
		$^{208}$Pb  &  0.74     & 0.14      &  0.14     & 0.20   & 0.37    & 0.60     & 0.76    \\  \hline
		(Q)PVC &  & & &  &&& \\
		$^{48}$Ca   &  0.70     & 0.25      &  0.36     & 0.51   & 0.67    & 0.90     & 1.07     \\
		$^{120}$Sn  &  0.67     & 0.14      &  0.02     & 0.18   & 0.36    & 0.68     & 0.82    \\ 
		$^{208}$Pb  &  0.94     & 0.37      &  0.25     & 0.06   & 0.08    & 0.31     & 0.48   \\  \hline \hline   
	\end{tabular}
	\caption{The deviation of ISGMR energies from experimental data [$|E_c^\textnormal{theo.}- E_c^\textnormal{exp.}|$ (MeV)] in $^{48}$Ca, $^{120}$Sn, and $^{208}$Pb, calculated by (Q)RPA and (Q)PVC using the Skyrme parameter sets SkP, SkM*, SV-K226, KDE0, SV-bas, SV-K241, and SAMi. The experimental data are taken from  \cite{T.Li_2007_PRL, Patel_2013_PLB, Olorunfunmi_2022_PRC}.}	\label{tab-1}
\end{table}

In summary, we have developed a fully self-consistent approach, that is, a Quasiparticle Random Phase Approximation (QRPA) plus Quasiparticle-Vibration Coupling (QPVC) model, based on Skyrme-Hartree-Fock-Bogoliubov. We have clearly demonstrated that the inclusion of QPVC effects is crucial in order to achieve a unified description of the monopole resonance in Ca, Sn, and Pb isotopes at the same time. 
The so-called softness of Sn isotopes is explained then, to a large extent, by the effects induced by QPVC. SV-K226 and KDE0 are found to give the best description of the experimental strength functions. We should add that (Q)PVC only accounts for the coupling to 2p-2h (or 4qp) configurations, and more complex configurations are missing. A theory that goes beyond may even further improve the agreement with experimental data in different mass regions.

Z.L. acknowledges the support by China Scholarship Council (CSC) under Grant No. 202106180051. This research was partly supported by the ¡°Young Scientist Scheme¡± of National Key Research and Development (R\&D) Program under Grant No. 2021YFA1601500 and Natural Science Foundation of China under Grant No.12075104.

\end{document}